\documentclass[conference]{IEEEtran}
\usepackage[dvips]{graphicx}
\usepackage{cite}

\usepackage{url}
\begin{document}
\title{SAMI: Service-Based Arbitrated Multi-Tier Infrastructure for Mobile Cloud Computing}

\author{\IEEEauthorblockN{Zohreh Sanaei$^1$, Saeid Abolfazli$^2$, Abdullah Gani$^3$, Muhammad Shiraz$^4$}
\IEEEauthorblockA{Mobile Cloud Computing Research Lab\\
Faculty of Computer Science and Information Technology\\
University of Malaya, Kuala Lumpur, Malaysia \\
Email: sanaei$^1$, abolfazli$^2$, abdullahgani$^3$@ieee.org,  muh\_shiraz@siswa.um.edu.my$^4$}
}
\maketitle

\begin{abstract}
Mobile Cloud Computing (MCC) is the state-of-the-art mobile computing technology aims to alleviate resource poverty of mobile devices. Recently, several approaches and techniques have been proposed to augment mobile devices by leveraging cloud computing. However, long-WAN latency and trust are still two major issues in MCC that hinder its vision. In this paper, we analyze MCC and discuss its issues. We leverage Service Oriented Architecture (SOA) to propose an arbitrated multi-tier infrastructure model named SAMI for MCC. Our architecture consists of three major layers, namely SOA, arbitrator, and infrastructure. The main strength of this architecture is in its multi-tier infrastructure layer which leverages infrastructures from three main sources of Clouds, Mobile Network Operators (MNOs), and MNOs' authorized dealers. On top of the infrastructure layer, an arbitrator layer is designed to classify Services and allocate them the suitable resources based on several metrics such as resource requirement, latency and security. Utilizing SAMI facilitate development and deployment of service-based platform-neutral mobile applications. 
\end{abstract}

\IEEEpeerreviewmaketitle

\section{Introduction}
\IEEEPARstart Recently, cloud computing technology has created an impetus ground for augmenting mobile devices and born mobile cloud computing (MCC) paradigm in order to tackle mobile devices' resource constraints (e.g. storage capacity, computational resources, and battery power). MCC is more advanced than mobile computing and cloud computing due to inheritance of mobility attribute from the former and resource richness from the latter. This advantage will provide a vast opportunity for different mobile users to have a ubiquitous computing regardless of location, time, and type of systems (e.g. smartphones, tablets, and notebooks). 

Recent study by ABI Research forecasts that the number of MCC subscriber worldwide will be almost 1 billion by 2014 \cite{ABIsub}. Also, driving nearly \$5 billion in revenues by more than 240 million mobile business users by the end of 2015 \cite{ABIResearch}. In our recent study \cite{ZohrehSanaei2012}, we connoted that MCC environment requires ubiquitous, energy efficient, and trustworthy approaches, systems, and architectures to be flourished. Moreover, we reviewed several mobile augmentation approaches \cite{SaeidAbolfazli2012} like offloading applications to distance resource-rich devices \cite{Chun2011} or proximate devices \cite{Satyanarayanan2009}, and developing lightweight applications like \cite{March2011}, but solution that can cover all the aforementioned MCC requirements is lacking.

Furthermore, data trafficking will be drastically increased by Internet-based cloud services for cloud-mobile users based on a recent report by Informa Telecoms \& Media \cite{internettraffic}. The annual data upload and download rates by end of 2015 are expected to grow as much as seven times compare with the same in 2010. Such significant growth is due to popularity of attractive internet-based services like BBC iPlayer \cite{iPlayer}, Apple-iCloud \cite{icloud}, and Dropbox \cite{dropbox} as a cloud-based online storage. Besides, another report by Morgan Stanley \cite{Morgan2009} shows, MNOs’ data revenues are not succeeding the rise as much as data traffic. Data traffic from 2008 to 2013 is changing to 76\%, while the same time data revenues just growing 13\%. 

Therefore, MNO's reduced workload creates an opportunity to leverage their resources and infrastructures and propose a MCC architecture to benefit larger mobile community. This MNO-friendly architecture will create further business opportunity for MNOs. we are endeavouring to propose a novel MCC architecture based on service oriented architecture (SOA) \cite{Huhns2005} by leveraging resources of the giant cloud service providers (like Amazon and Google) and infrastructures of MNOs. This paper attempts to establish a loosely coupled interoperation between mobile users, cloud providers, and MNOs and their authorized dealers.

Rest of the paper is organized as follows: Section II defines and analyzes MCC and discusses its major challenges. Section III describes multiple roles of MNOs. we propose the SAMI (Service-Based Arbitrated Multi-Tier Infrastructure) architecture for Mobile Cloud Computing in section IV. Section V presents benefits and drawbacks of the architecture and Section VI concludes this paper.

\section{Mobile Cloud Computing}
MCC is an amalgam of three foundations, namely cloud computing, mobile computing, and networking \cite{ZohrehSanaei2012}. The most promising and intriguing characteristics of MCC paradigm are mobility and rich functionality. We define mobile cloud computing as ``\textit{a rich mobile computing technology that leverages unified elastic resources of varied clouds and network technologies toward unrestricted functionality, storage, and mobility. It serves a multitude of mobile devices anywhere, anytime through the channel of Ethernet or Internet regardless of heterogeneous environments and platforms based on the pay-as-you-use principle}''.

MCC exploits advantages of distributed, distance, resources-rich cloud infrastructures and has been conceived to augment a multitude of mobile devices, especially smartphones and alleviate their shortcomings. The futuristic accomplishments of MCC are envisioned to be deployed in several areas like healthcare (e.g. telemonitoring and telesurgery), education, IT business, rural and urban development, and social networking. In this regard, several efforts such as \cite{cuervo2010maui,Zhang2011a,Chun2009,March2011,Lu2011,Kemp2010a} deploy cloud computing technology to enhance the capability of smartphone. Moreover, leveraging cyber-resources is beneficial to enhance information safety on smartphones since they are susceptible to theft, loss, and physical damage. DropBox and SugarSync \cite{sugarsync} as cloud storage services are able to synchronize data in different computing devices (e.g. smartphones, Tablets, and laptops). However, its adoption is inhibited by several issues that are explained as follows.
 
\indent $\bullet$ \textit{\textbf{Seamless Connectivity:}} Continuous and consistence communication is the most important criterion in MCC that lays on further research and innovative mechanisms. Wireless networks have low-bandwidth, intermittent, and less-reliable transmission grounds compared with the wired networks. Dismissal of always-on connectivity, excessive consumption of limited mobile resources, and disproportionate delaying of application execution are some of the most important problems that cause QoS computing degradation. Also, reliable inter-system and intra-system signal handoff schemes beside inter-connectivity across heterogeneous wireless networks and wired networks are fundamental characteristics to reach a seamless ubiquity in a dynamically changing environment. However, to address some of these issues, next generation wireless networks \cite{Nasser2006} and the patented open mobile infrastructure with open network architecture and open wireless architecture (OWA) by Sieneon \cite{sieneon}, define the future solutions for MCC.

\indent $\bullet$ \textit{\textbf{Long WAN Latency:}} Latency is a fundamental obstacle that confines MCC solutions to a low-latency approaches. Energy efficiency \cite{Miettinen2010} and interactive response-time \cite{Lagar-Cavilla} are two major metrics in remote processing applications which are adversely impacted by latency in MCC landscape. To reduce the long WAN latency, the idea of utilizing WLAN instead of HSDPA (High Speed Download Packet Access) to process heavy functionalities in nearby computing systems called ``cloudlet'' has been proposed \cite{Satyanarayanan2009}. However, trust and security issues as the most important criteria prevent the offered solution to cover user's confidence with the cloudlet out-source infrastructure. Hence, further research advancements is required in both to develop more trustworthy systems and crisper response approaches. Moreover, the recent study by Nokia Research \cite{Miettinen2010} depicts that data transfer bit-rate imposes comparatively more impact on the energy efficiency of cellular networks than WLAN. The higher the transmission bit-rate, the more energy efficient the transmission. Hence, we can summarize that, the more dynamic the environment, the more intelligent and scalable context-aware systems in MCC.

\indent $\bullet$ \textit{\textbf{Cloud's trustworthiness:}} Trust is an essential factor for the success of the burgeoning MCC paradigm \cite{ZohrehSanaei2012} while its construction in a secure manner within a non-uniform platform (e.g. heterogeneous infrastructures and networks, fragmented mobile devices and servers) is a challenging task. Furthermore, leveraging the cloud services by crossing the channel of Internet as a bridge between cloud consumers and cloud providers (over wireless and wired networks) exacerbates the trust issue with existence of malware applications. Hence, ensuring users about their data confidentiality, information security and privacy, and data integrity are crucial issues, because neither cloud-users have physically access to remote systems, nor know about the real business policy of each cloud service providers. To cope with these problems we list some of the most important approaches from \cite{Satyanarayanan2009},\cite{trustworthy} as follows.

\begin{figure*}[!ht]
\begin{center}
\includegraphics[width=5in, height=4.0in]{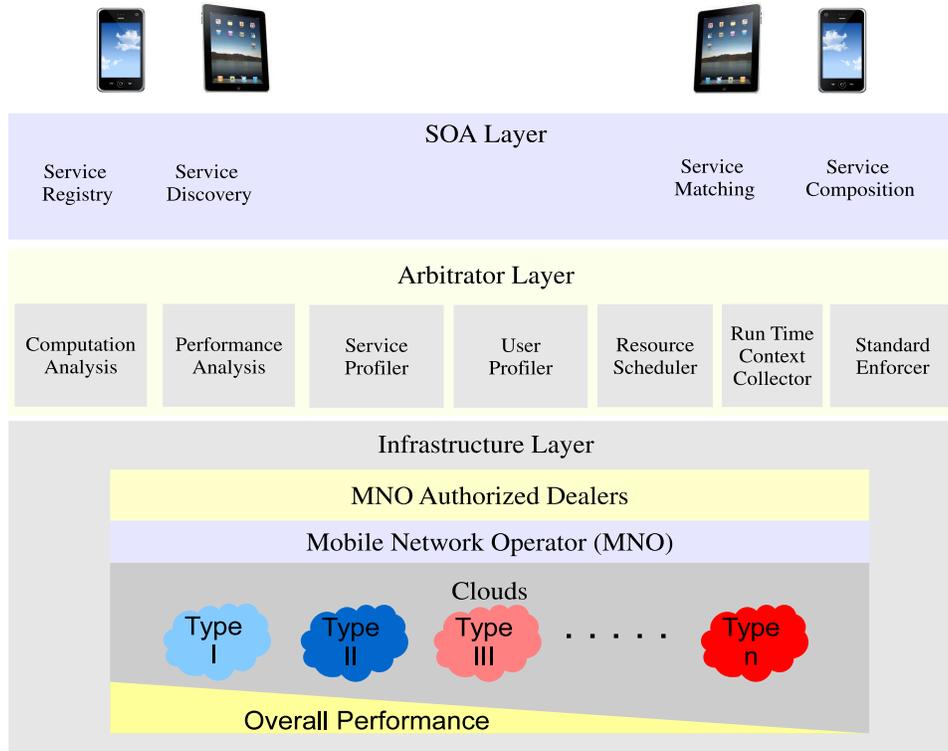} 
\caption{SAMI: The Service-Based Arbitrated Multi-Tier Infrastructure for Mobile Cloud Computing} \label{architecture}
\end{center}
\end{figure*}
\begin{itemize}

\item \textit{Trust Establishment:} User can pre-use and test the host systems before trust the service providers which is more protective and robust but also more cumbrous. 

\item \textit{Trust Aggregation:} Each node can trust to another node based on trust aggregating from its nearby node or trust-aggregator server. This approach may ease the way to trust to another node but aggregating trust is a complex and time consuming task which might not be energy deficient for mobile devices due to delays.  

\item  \textit{Indirect Trust:} Node X trust to node y, and Node Y has already trusted to node Z, hence, node X will trust to node Z. This approach is quick and logically true, but level of trust in middle nodes is not clear for other nodes. Therefore, it is useful for jobs that need low-level of security 

\item \textit{Trust Based on Reputation:} A node(consumer) can trust to another node (service provider) based on the legal business of service provider's identity or some other external considerations that show how much the service provider is optimistically accountable and reliable. Although this approach is fast, but is more vulnerable. 
\end{itemize}
\indent $\bullet$ \textit{\textbf{MCC Billing:}} MCC is landed in the rapidly changing environment that distinguishes its billing system from cloud computing. Long WAN delays, jitter, session re-establishment delay, bandwidth capacity, and security degree are major parameters that diverge MCC billing system form systems settled in the static environment. Utilizing MNOs as a bridge between cloud-mobile users and cloud providers can facilitate the MCC billing design due to pre-established of the fundamental roles in mobile computing environment. 

\indent $\bullet$ \textit{\textbf{Heterogeneity in MCC:}} Heterogeneity is a big challenge that covered MCC ground and raised many discrepancies. Interoperability between fragmented systems with varied APIs, data integrity within/across heterogeneous data warehouses, and migrating data/code from one cloud to another cloud and from mobile devices to the cloud or in reverse are highlighted issues that create new research area.

\section{MNOs with Multiple Roles}
MNOs are well-established, reputed, and robust infrastructure providers that have served mobile users since the beginning of mobile communication. One of the requirements in our architecture is to have a robust billing operation which is already available on MNOs. Hence, they are an appropriate candidate to be chosen to play multiple roles in this architecture. We briefly describe MNO's Roles as follows.

\begin{itemize}

\item{\textit{MNO as a cloud provider:}} MNOs can provide storage as a service(SaaS), infrastructure as a service (IaaS), and software as a service (SaaS) to their end users by deploying Cloud concept. AT\&T is the first MNO that offers cloud computational and storage infrastructure to their clients \cite{ATT1}.

\item{\textit{MNO as a cloud broker:}} MNOs can play as service mediator between cloud providers such as Azure and Google and mobile users, and from the other aspect, being a broker across other MNOs. In order to increase efficiency, MNOs can interoperate to each other to serve larger community of mobile users.

\item{\textit{MNO as a cloud consumer:}} MNOs can utilize cloud vendors' services if their proprietary resources are not enough to serve their clients. Hence, MNOs leverage cloud services like IaaS from IT giants' cloud service providers based on a negotiated service level agreement (SLA).

\item{\textit{MNO as a carrier cloud:}} Carrier cloud MNOs meet the needs for connectivity, reliability, and accountability between heterogeneous wireless networks and wired systems in the cloud which are critical issues in MCC domain. 
\end{itemize}

\section{SAMI: Service-Based Arbitrated Multi-Tier Infrastructure}
We propose an architecture to tackle aforementioned issues that are mainly created due to heterogeneity and trust which is depicted in Figure  \ref{architecture}. We leverage SOA since it is a service-driven approach to generate platform-neutral applications. Our proposal consists of three major layers, namely SOA, arbitrator, and infrastructure. The main strength of this architecture is in its multi-tier infrastructure layer which leverages infrastructures from three main sources of Clouds, MNOs, and MNOs' authorized dealers. On top of the infrastructure layer, arbitrator layer is responsible to classify Services and allocate the suitable resources to them based on several metrics such as resource requirement, latency, and security. We describe each layer and their respective components as follows.

\subsection{SOA Layer} 
Service oriented layer is responsible to perform Service-related tasks such as Service registry, discovery, and composition that are explained as follows.

\begin{itemize}
\item Service Registry: Service registry is a functional component that receives service registry requests from service developers (who has implemented the Service) and maintains relevant data in a local database. When a new Service is registered in the system, it initiates a request to another component called resource scheduler in lower layer (arbitrator layer) to allocate a resource according to the nature of the Service. The allocated resource will be stored in the service registry database for future reference. When a Service consumer sends a discovery message, the Service registry browses the database and sends a message containing the address of Service provider by which service requester can utilize the Service.

\item Service Discovery: Service discovery is one of the runtime processes responsible to reply to a SOAP message received from the client application to find a specific Service. The found Service specification will be forwarded through another SOAP message to the sender.

\item Service Matching: Service matching is an alternative method to find a Service when there is no specific information about the Service name or its address. Based on certain specification about Service functionality presented in description section of registered Services, the Service can be found.

\item Service Composition: If the required functionality is not presented or defined by a single Service, the Service composition can create a composition of some already-developed, basic Services to accommodate the desired functionality. Service composition is crucial in this architecture to extend reusability of Services.

\end{itemize}
\subsection{Arbitrator Layer} 
In this layer MNOs act as arbitrators between front-end (Cloud-mobile users) and back-end (Cloud service providers). It receives resource allocation requests from SOA layer and monitors infrastructure layer to determine where the Service should be allocated for more efficient execution. One decided, a stand-alone copy of the Service will be transferred to the infrastructure for future reference. In this way, the network overhead is noticeably reduced. 

\begin{itemize}
\item Resource Scheduler: When a new Service is registered and stored in the Service registry, the allocation request is sent to the resource scheduler. The resource scheduler analyzes the intrinsic and descriptive properties of the Service and determines where the Service should be executed to serve its clients more efficiently. Figure \ref{flowchart} depicts a simple decision making process by Service scheduler.  

\begin{figure}[!ht]
\begin{center}
\includegraphics[scale=0.5]{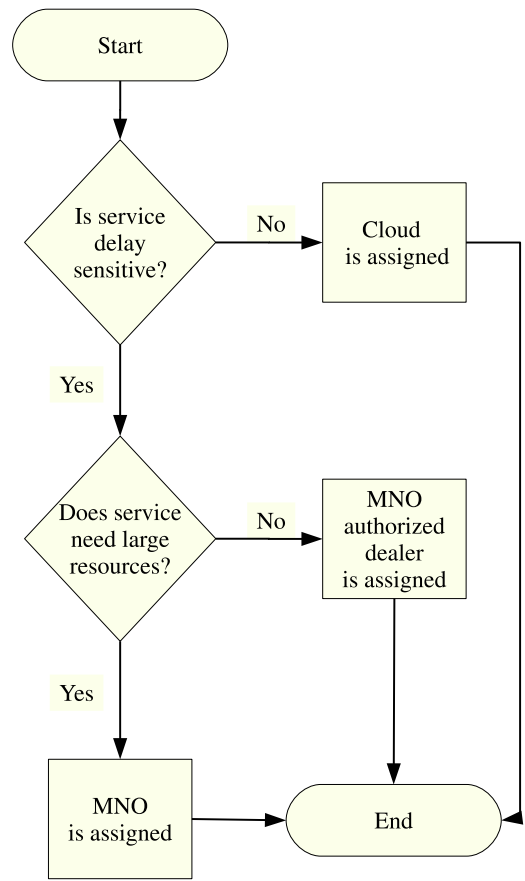} 
\caption{A generic decision making flow of resource scheduler} \label{flowchart}
\end{center}
\end{figure}

\item Service Profiler: One of the broker's responsibilities is to monitor validity and performance of Services and substitute them with more appropriate Services if required. Service profiler component is defined to evaluate the functionality and performance of Services in real scenario. If the Service functionality is not in agreement with the Service description, then the Service will be replaced by a more efficient Service to keep the system efficient.

\item User Profiler: Since the Services are invoked by different Service consumers, their organizational or individual preferences are important. Therefore, user profiler component is responsible to monitor and store users preference to be able to customize the service delivery and execution based on their criteria.

\item Run Time Context Collector: In order to increase system efficiency and performance, a runtime context gathering is a vital process. Such context information can be leveraged to reschedule resources in our architecture.

\item Standard Enforcer: Standardization is one of the heterogeneity handling techniques which is useful to be deployed in heterogeneous MCC \cite{ZohrehSanaei2012}. Hence, considering standards in designing our architecture is important. This component is responsible to monitor and enforce standardization for Services.

\item Computational Analysis: 
During initial steps of Service registry, a certain infrastructure layer is assigned to each Service. However, it might not be the optimal choice. Hence, to achieve the optimal performance, computational analysis component monitors computational capabilities of assigned resources. If locating the Service to another layer enhances the quality of execution, the resource scheduling will be called for re-scheduling the resources. 

\item Performance Analysis:
Since our architecture is aimed to alleviate network latency, performance analysis component is deemed to analyze communication latency and resource consumption pattern of different Service. If the amount of resources required to run a particular Service is not in balance with other factors such as communication delay, the resource scheduling tasks should be repeated to determine more appropriate infrastructure. Additionally, the relation between frequency of a Service invoking and execution and the amount of resources consumed is monitored by this component. The result of this analysis is useful to enhance the quality of overall execution. For example, if a delay-sensitive Cloud Service is invoked too frequently, the Service can be moved to a nearby MNO or its authorized dealer to tackle the communication latency.
\end{itemize}

\subsection{Infrastructure Layer}
In order to enhance quality of computing, decrease communication latency, and improve energy efficiency we propose a multi-tier infrastructure layer including three major layers, namely clouds, MNOs, and MNO's authorized dealers which are explained as follows. 

\begin{itemize}
\item{\textit{MNO Authorized Dealers:}}
MNO authorized dealers are those service outlets that are scattered in this urban or rural areas to directly communicate with MNO's clients to serve them better. The MNO's usually consider certain criteria when offering dealership and make them responsible against certain behavioural disciplines. Such assumption isolates and promotes them from other shops and service outlets. The closest infrastructure layer to the mobile Service consumer are MNO's authorized dealers that can play the role of trusted surrogates which are able to serve mobile users in vicinity. Large number of authorized dealers are mainly scattered in different business spots such as shopping mall, market areas, airports, and commercial buildings where plenty of mobile users are operating. Currently Maxis, one of the largest Malaysian MNOs, has 1372 authorized dealers \cite{Maxisdealer} which is noticeable number of outlets each with couple of wall-connected power source and wired network connection. Utilizing computing infrastructures of these nearby resources is deemed not only to decrease communication latency and execution time, but also to increase computing capability of mobile devices. However, accessing their resources is limited mostly to the operating hours and weekdays and their computing resources (ie. CPU, memory and storage) are limited. Therefore, they can only serve specific group of clients. Usually latency-sensitive Services are expected to be executed on these machines due to lower communication latency. Nevertheless, futuristic business opportunity and resource renting can motivate MNO's authorized dealers to increase their local resources.

\item{\textit{Mobile Network Operator (MNO):}}
To complement MNO authorized dealers, MNOs as a well-established, reputed business stakeholder can play an important role in our architecture. MNOs are able to provide continuous services to their client without need to access to the Internet. One of the vital properties of MNOs is their ability to communicate using cellular network without entering the insecure channel of Internet, which increases computing security. Giant telecommunications service providers such as AT\&T have recently started to offer computational Services as Cloud to their mobile users \cite{att}, which is a proof-of-concept to our multi-tier architecture. 

\item{\textit{Clouds:}}
The third layer in our architecture includes Clouds infrastructures offered by giant IT companies such as Amazon S3 \cite{amazon} and Google App Engine \cite{googleappengine}. Though these resources are often located in distance and user should go through the channel of Internet, they can offer elastic resources equal to the Service requirements without upfront resource provisioning. Data-intensive services often utilize Cloud infrastructures due to large amount of data. In order to obtain optimal arbitration, Cloud resources are classified based on their overall performance. Some of the quality metrics in this classification are execution latency, communication bandwidth, service cost and billing, and security. The red color Cloud (far right) in Figure \ref{architecture} indicates lowest degree of performance while the light blue Cloud (far left) represents a high performance Cloud.
\end{itemize}

\section{Advantages and Disadvantages}
In this section we describe benefits and limitations of our architecture.
\subsection{Advantages}
\textbf{Portability:} This architecture utilizes the concept of SOA which is a widely-accepted solution to overcome application portability in MCC. Mobile applications are adversely affected by the current platform, hardware, feature, and API heterogeneity of mobile devices. Therefore, an application cannot be ported from one mobile device to another non-similar device which increases development cost and time for mobile applications. To alleviate portability, we employed SOA to enable development of platform-agnostic mobile applications. 

\textbf{Energy Dissipation-Prevention:} Energy dissipation is prevented from three perspectives by the proposed architecture: (1) mobile users are able to utilize the closest available resources to bring down the communication latency. Hence, the application execution time is reduced and energy efficiency is increased. (2) Surrogate machines in earlier approaches -due to their non-responsibility and stochastic situation- are able to discontinue the job with mobile devices which leads to energy-squandering that is addressed in our architecture (3) There are globally many distributed servers in MNOs and their branches which are not solely profitable but together can establish a Cloud and provide cloud services toward saving energy and hiking revenue. 

\textbf{Automatic Resource Discovery:} One of the most irksome impediments of mobile users is to identify the most suitable resource for their computational requirements. The arbitrator layer in SAMI, takes this headache away from mobile users and ensure them the optimal resource.

\textbf{Trustworthy:} Based on trust approaches that explained in section II, placing MNOs as an intermediate layer between mobile devices and cloud systems is more secure than previous models like cloudlet, cyber foraging \cite{Satyanarayanan2001} approaches, and Hyrax \cite{Marinelli2009} that consume surrogate machines.

\textbf{Interoperability:} MNOs not only in heterogeneous wireless networks acts as a well-established carriers, but also can act as an adaptor in two different edges; wired (connectible to the other cloud like Google) and wireless(connectible to mobile devices). Therefore, this multi-adapter can facilitate data interoperation in MCC.

\textbf{MNOs' Revenue Hike:} Several roles of MNOs in SAMI architecture enable them to boost their enterprise and business opportunities. MNOs not only have high potential and chance to become a widely-trusted cloud provider for mobile devices, but their authorized dealers can benefit from this architecture.

\subsection {Disadvantages}

\textbf{MNO Overhead:} This architecture imposes a processing overhead on MNOs (because they play the role of arbitrator) due to continuous arbitration process. For every Service invoking the Service profiler and run time context collector components should crawl to gather essential data.

\textbf{Complex Management:} Our decentralized architecture which works on three different infrastructural layers imposes extra management and maintenance cost.

\section{Conclusions}
Mobile cloud computing is the key technology to alleviate two major mobile computing shortcomings, namely resource poverty and battery limitation of mobile devices. Storing data and executing heavy functionalities on remote computing rich resources are the most promising and intriguing characteristics of MCC. However, heterogeneity, long WAN latency, trustworthy, reliability, and lack of suitable billing systems in MCC are required to be addressed. Moreover, MCC landscape is a significant opportunity for MNOs to play a vital role like cloud provider and cloud broker to raise their profit. 

This paper proposed the SAMI: Service-based Arbitrated Multi-tier Infrastructure for mobile cloud computing that allows mobile users to access the most proximate resources with least latency and higher trust. Due to unique characteristics of MNOs (i.e. well-established communication entities in heterogeneous mobile computing environment, mobile billing systems, and high degree of trust reputation) they are the best candidate to serve as an arbitrator in our SAMI architecture. More work will be conducted on this research to implement and validate the SAMI architecture.

\section*{Acknowledgment}
This work is funded by the Malaysian Ministry of Higher Education under the University of Malaya High Impact Research Grant UM.C/HIR/MOHE/FCSIT/03.

\bibliographystyle{IEEEtran}

%\bibliography{D:/zohreh/Dropbox/Jabref/ieeefull}
\bibliography{ieeefull}
\end{document}